\documentclass[prb,twocolumn,a4paper,showpacs,showkeys]{revtex4}
\usepackage{graphicx}
\usepackage{amsmath}
\usepackage{amssymb}
\usepackage{dcolumn}
\newcommand{\me}{\text{e}}
\renewcommand{\vec}[1]{\mbox{\boldmath$#1$}}

\newcommand{\dif}{\mathrm{d}}

\begin{document}
\title{Intersubband carrier scattering in $n$- and $p$-Si/SiGe quantum
wells with diffuse interfaces}
\author{A. Valavanis}
\email{a.valavanis05@leeds.ac.uk}
\author{Z. Ikoni\'{c}}
\author{R. W. Kelsall}
\affiliation{Institute of Microwaves and Photonics, School of Electronic
and Electrical Engineering, University of Leeds, Leeds LS2 9JT, United
Kingdom}
\date{\today}

\begin{abstract}
Scattering rate calculations in two-dimensional Si/Si$_{1-x}$Ge$_x$
systems have typically been restricted to rectangular Ge profiles at
interfaces between layers. Real interfaces however, may exhibit diffuse
Ge profiles either by design or as a limitation of the growth process.
It is shown here that alloy disorder scattering dramatically increases
with Ge interdiffusion in (100) and (111) $n$-type quantum wells, but
remains almost constant in (100) $p$-type heterostructures. It is also
shown that smoothing of the confining potential leads to large changes
in subband energies and scattering rates and a method is presented for
calculating growth process tolerances.
\end{abstract}

\maketitle

\section{Introduction}
Two dimensional intersubband devices in the Si/Si$_{1-x}$Ge$_x$
materials system offer a possible reduction in fabrication costs
compared with more conventional III--V systems.\cite{book:Pavesi2003}
Successful operation of resonant tunneling diodes (RTDs) has been
achievable for several years,\cite{article:APLLiu1988} and
electroluminescence from more complex quantum cascade structures has
been observed.\cite{article:ScienceDehlinger2000, article:APLBates2003}
A quantum cascade laser (QCL) has not yet been developed in 
Si/Si$_{1-x}$Ge$_x$ although several designs have been proposed
recently.\cite{article:APLDriscoll2006, article:SSTHan2007}

In order to accurately design and simulate such structures, a good
understanding of intersubband carrier dynamics is required. Previous
models have assumed that interfaces between layers are perfectly
abrupt,\cite{article:JAPIkonic2004} while in reality, diffuse Ge
profiles may result either by design, by interdiffusion during the
growth process or by surface segregation of Ge
atoms.\cite{article:SurfSciZhang2006}

The Ge interdiffusion dramatically changes the subband
spacing\cite{article:IEEELi1996} and the overlap between wave functions.
A more accurate model of intersubband scattering rates must therefore
account for these effects. In this paper, we review the models of the
principal intersubband scattering rates and extend the conventional
interface roughness scattering model to an arbitrary interface geometry.
To determine the effect on simple intersubband systems, scattering rates
were calculated as a function of subband spacing, electron temperature,
and diffusion length in single quantum wells (QWs) in (100) $p$-type and
(100) and (111) $n$-type systems.

Knowing the effect of interdiffusion on scattering rate, it is possible
to estimate the robustness of device designs. By calculating
intersubband scattering rates and subband spacings, design tolerance to
interdiffusion may be estimated. The viability of a design may be
assessed by comparing this with the capabilities of growth processes. We
use the example of a coupled QW system to demonstrate the technique and
predict the level of interdiffusion required to cause device failure.

\section{Scattering models}
Carrier scattering in a two dimensional Si/Si$_{1-x}$Ge$_x$ system may
be described by a set of independent processes. The models for the
Coulombic, electron--phonon, and alloy disorder interactions apply to
arbitrary interface geometries, while some modification is required for
interface roughness scattering. The mechanisms are summarized as
follows.

\subsection{Interface roughness scattering}
In $z$-confined two-dimensional heterostructures, the confining
potential varies with fluctuations in interface location over the
$(x,y)$ plane.\cite{article:PRPrange1968,article:SSTPenner1998} The
roughness is usually assumed to have a Gaussian Fourier transform
$\Delta_z(\vec{r})$ with height $\Delta$ and correlation length
$\Lambda$, which is isotropic across the $(x,y)$
plane,\cite{article:APLSakaki1987, article:APLTsujino2005,
article:RMPAndo1982} such that
\begin{equation}
 \left\langle\Delta_z(\vec{r}) \Delta_z(\vec{r'})\right\rangle =
 \Delta^2\exp{\left(-\frac{|\vec{r}-\vec{r'}|^2}{\Lambda^2}\right)}.
\end{equation}

The commonly used expression for the resulting scattering 
rate\cite{article:JAPUnuma2003} assumes an abrupt interface geometry.  
This has been accurately fitted to experimental data for structures with
approximately abrupt interfaces,\cite{article:PRBCalifano2007} but the 
expression is incompatible with smooth envelope potentials.  We 
determine the scattering rate for an arbitrary interface geometry and 
verify that it reduces to the specific case of an abrupt interface.

The perturbation $\Delta_V(\vec{R})$ due to a position shift
$\Delta_z(\vec{r})$ in an arbitrary confining potential $V(z)$ is 
assumed to be correlated over the length of a single interface.  At the 
point $\vec{r}=\vec{x}+\vec{y}$ assuming isotropy across the $xy$ plane,
\begin{equation}
\label{eqn:Perturbation}
 \Delta_V(\vec{R})=V[z-\Delta_z(r)]-V(z)\approx-\Delta_z(\vec{r})
 \frac{\dif{V(z)}}{\dif{z}}.
\end{equation}

If the $I$-th interface in a multilayer structure is centered about the 
plane $z=z_I$ and extends over the range $(z_{L,I},z_{U,I})$, we define
the scattering matrix element as
\begin{equation}
 \label{eqn:IRMatrixElement}
 F_{fi,I}=\left\langle{}f\left|\frac{\dif{V}}{\dif{z}}
 \operatorname{rect}\left(\frac{z-z_I}{z_{U,I}-z_{L,I}}\right)
 \right|i\right\rangle,
\end{equation}
where $\left|f\right\rangle$ and $\left|i\right\rangle$ are the final 
and initial wave functions respectively and the rectangular function
$\operatorname{rect}(z)$ is defined as
\begin{equation}
\operatorname{rect}(z) = \left\{
\begin{array}{cc}
1;& |z| \leq 0.5\\
0;& |z| > 0.5
\end{array}\right..
\end{equation}

From Fermi's golden rule, the scattering rate $W_{fi,I}(k_i)$ 
is\cite{article:JAPUnuma2003}
\begin{equation}
\label{eqn:IRRate}
 W_{fi,I}(k_i) = \frac{m_d\Delta^2\Lambda^2|F_{fi,I}|^2}{\hbar^3}
 \int_0^\pi\mathrm{d}\theta\,\me^{-k_{fi}^2\Lambda^2/4},
\end{equation}
where $\vec{k_{fi}}=\vec{k_f}-\vec{k_i}$ is the scattering vector,
$\vec{k_f}$ and $\vec{k_i}$ are the final and initial wave vectors
respectively, $\theta$ is the scattering angle and $m_d$ is the
density-of-states effective mass.

At this stage, the general result may be tested against the example of 
an abrupt change in envelope potential of magnitude $V_{0,I}$, where
the perturbating potential is
\begin{equation}
 \Delta_V(\vec{R})=V_0(z_I)\operatorname{rect}\left\{
  \frac{1}{\Delta_z(\vec{r})}
  \left[z-\left(\frac{\Delta_z(\vec{r})}{2}+z_I\right)\right]
 \right\}.
\end{equation}
Under time-independent perturbation theory, the perturbation must be 
small, \emph{i.e.} $\Delta_z(\vec{r})\to{}0$ and the perturbing
potential simplifies to
\begin{equation}
 \label{eqn:PerturbationAbrupt}
 \Delta_V(\vec{R})=V_0(z_I) \Delta_z(\vec{r}) \delta\left(z-z_I\right).
\end{equation}
The scattering matrix element (Eqn.\ \ref{eqn:IRMatrixElement}) becomes
\begin{equation}
 F_{fi,I}=\left\langle{}f|V_{0,I}\delta(z-z_I)|i\right\rangle,
\end{equation}
in agreement with the well known expression.\cite{article:JAPUnuma2003}

Returning now to the general expression for scattering rate 
(Eqn.\ \ref{eqn:IRRate}), we determine the total scattering rate for a 
structure with $N$ layers, numbered $I\in\mathbb{N}^+, I\leq{}N$. The 
integral over $\theta$ simplifies to a regular modified cylindrical
Bessel function of zeroth order, $I_0$.\cite{book:Gradshteyn2000}  By
assuming roughness profiles are uncorrelated over separate interfaces, 
the total rate is found as a summation,
\begin{equation}
\label{eqn:IRRateSmooth}
 W_{fi}(k_i)=B_{fi}(k_i)I_0\left(\frac{k_fk_i\Lambda^2}{2}\right)
 \Theta(\alpha^2),
\end{equation}
where the Heaviside step function, $\Theta$ permits a nonzero rate only
for real final wave vectors. The new matrix element is
\begin{equation}
 B_{fi}(k_i)=\frac{m_d\pi}{\hbar^3} \left(\Delta\Lambda\right)^2 
 \sum\limits_{I=1}^{N-1}   |F_{fi,I}|^2\me^{
  -\frac{\Lambda^2}{4} \left(k_i^2+\alpha^2\right)
 },
\end{equation}
where $\alpha=\sqrt{k_i^2-\frac{2m_d}{\hbar^2}E_{fi}}$.

The new model was fitted to the experimental data described in detail in
our previous paper.\cite{article:PRBCalifano2007} The parameters,
\{$\Delta$=1.4\,\AA, $\Lambda$=50\,\AA\} accurately fit the measurements
and are very similar to our previous theoretical
values,\cite{article:PRBCalifano2007} and to other recent
data.\cite{article:PhysStatSolHuan2007, article:APLTsujino2005} The
slight difference in fitting parameters arises from numerical 
approximations in the perturbing potentials: the perturbation for 
arbitrary interface geometries (Eqn.\ \ref{eqn:Perturbation}) uses a 
Taylor series expansion, whereas the solution for abrupt interfaces
(Eqn.\ \ref{eqn:PerturbationAbrupt}) uses the Dirac $\delta$ function
limit of the narrow rectangular function.

\subsection{Alloy disorder scattering}
A standard scattering model for alloy disorder
scattering,\cite{article:PRBQuang2007, article:PRLMurphy-Armando2006}
has been modified slightly to permit variable alloy composition, $x(z)$.
The resulting scattering potential becomes
\begin{equation}
 \left|\left\langle{}U_{\text{AD}}(q)\right\rangle\right|^2=\frac{a_0^3
 \delta{}V^2}{8}\int{}\psi_f^2(z)x(z)[1-x(z)]\psi_i^2(z)\,\dif{}z,
\end{equation}
where $a_0$ is the in-plane lattice constant and $\delta{}V$ is commonly
approximated as the difference in conduction band potentials between Si
and Ge. As $x(1-x)=0$ for a pure Si layer, the integral domain in abrupt
interface systems is restricted to the barriers in $n$-type systems or
the wells in $p$-type systems. In a diffuse system however, the Ge
content is always nonzero and the integral domain extends over the
entire structure.

\subsection{Carrier--phonon scattering}
Wells in $p$-type heterostructures contain similar fractions of Si and
Ge. Hole--phonon scattering via the deformation potential interaction was
therefore calculated for the Si--Si, Si--Ge, and Ge--Ge branches of the
nonpolar optical mode.\cite{article:PRBIkonic2001,
article:JAPIkonic2004} In $n$-type heterostructures however, the Ge
fraction in the wells is small and only the Si--Si branch was
considered. Intravalley deformation potential scattering for the
acoustic mode was included for both $n$- and $p$-type heterostructures.

In $n$-type, Si-rich systems, scattering between conduction band
$\Delta$ valleys is described by either $g$ processes, which transfer
electrons to the opposite valley in reciprocal space or $f$ processes
which transfer electrons to the four perpendicular
valleys.\cite{article:PRBCanali1975} $g$-LO, $f$-LA, and $f$-TO
interactions are permitted in a zero-order model, whereas $g$-LA,
$g$-TA, and $f$-TA interactions are permitted only as first-order
processes \cite{article:PRBMonsef2002} and are somewhat slower. The
interactions are characterized by a deformation potential and a
frequency, $\omega_0$, which for intervalley interactions is nonzero.
The scattering rates increase rapidly with subband separation, $E_{fi}$
until they saturate at $E_{fi}\gtrsim\hbar\omega_0$.

\subsection{Coulombic interactions}
Ionized impurity scattering and carrier--carrier scattering rates were
calculated as Coulombic interactions between either a carrier and a
dopant ion or a pair of carriers. The expression for ionized impurity
scattering given by Unuma,\cite{article:JAPUnuma2003} was modified to
incorporate static screening in the Thomas-Fermi
approximation.\cite{book:Davies1998} Carrier--carrier scattering rates
were calculated using the screened Coulombic interaction model described
by Smet \textit{et.\ al}.\cite{article:JAPSmet1996} Both rates are fastest for
transitions between energetically similar states. Doping was set as
1$\times{}10^{16}$\,cm$^{-3}$ throughout each structure we considered,
as modulation doping is difficult to achieve in SiGe
heterostructures.\cite{article:SurfSciZhang2006} We also assume that all
dopants are ionized at low temperatures.  The sheet doping density was 
calculated as the product of the total volume doping and the length of
the structure.

\subsection{Average rates}
As justified previously,\cite{article:JAPJovanovic2006} all subband
electron temperatures were set to a single average value, $T_e$, assumed
to be different from the lattice temperature, which was taken as
$T$=4\,K in our calculations. The average scattering rate from the
second to first subband, $\overline{W}_{12}$ was calculated as
\begin{equation}
 \overline{W}_{12}=\frac{\int W_{12}f_2^\text{FD}(k_2)k_2\,\dif{}k_2}{\pi{}N_2},
\end{equation}
where intrasubband scattering was assumed to be much faster than
intersubband. The initial distribution of electrons is therefore an
equilibrium Fermi-Dirac function, $f_2^\text{FD}(k_2)$ using the
quasi-Fermi level for the subband, where $k_2$ is the initial wave
vector. The assumption has also been made that the destination states
are always unoccupied, which is reasonable at low doping levels.

\section{Diffuse quantum wells}
Annealing of an abrupt structure, with the Ge fraction $x_I$ in layer
$I$ provides a simple model of a diffuse system. The abrupt-interface
structure is embedded between infinitely thick barriers with composition
$x_0$. The composition profile after annealing
is\cite{article:IEEELi1996}
\begin{eqnarray}
x(z)&=&\frac{1}{2}\sum_{I=1}^N
x_I\left[\operatorname{erf}\left(\frac{z-z_{I-1}}{L}\right)-
\operatorname{erf}\left(\frac{z-z_I}{L}\right)\right]
\nonumber\\
&+&\frac{x_0}{2}\left[\operatorname{erf}\left(\frac{z-z_N}{L}\right)-\operatorname{erf}\left(\frac{z-z_0}{L}\right)\right],
\label{eq1}
\end{eqnarray}
where the $I$-th layer boundaries are $z_{I-1}$ and $z_I$, and $L$ is the diffusion length.

In the calculations presented below, we have assumed that the
composition profiles are symmetrical for the left and right interfaces
of a QW. This corresponds to the case where interdiffusion dominates
over surface segregation. If Ge segregation effects are strong, then the
interface profiles will be asymmetric, although the
effect on scattering is expected to be similar.

\begin{figure}
 \includegraphics{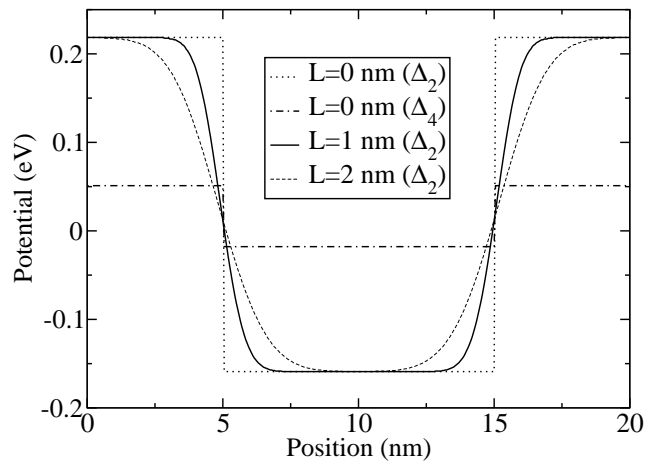}
 \caption{\label{fig:InterdiffN100}$\Delta_2$ conduction band edge for a
10\,nm Si QW between two 5\,nm Si$_{0.5}$Ge$_{0.5}$ barriers, with
varying Ge diffusion lengths. The $\Delta_4$ conduction band edge for
an abrupt interface is shown to be 120\,meV higher in energy.}
\end{figure}

A 10\,nm (100) $n$-type QW between two 5\,nm Si$_{0.5}$Ge$_{0.5}$
barriers is shown in fig.\ \ref{fig:InterdiffN100}, with the in-plane
lattice constant set to achieve strain balance. As $L$ increases, the
bottom of the well narrows and the top widens. The effect on scattering
in (100) $n$- and $p$-type and in (111) $n$-type systems is discussed in
depth in the following sections.

\begin{figure}
  \includegraphics{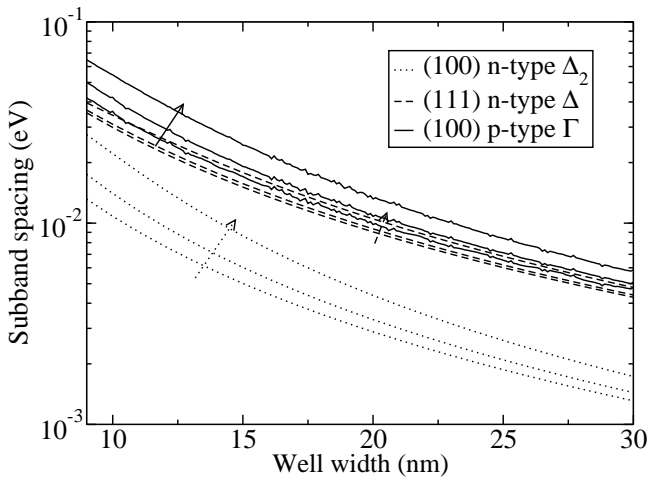}
  \caption{\label{fig:SubbandSpacing}Energy separation between the 
    lowest pair of subbands in a Si/Si$_{0.5}$Ge$_{0.5}$ QW as a 
    function of well width.  Results are shown for $\Delta$ valleys in
    $n$-type structures (Si wells) in the (100) and (111) orientations 
    and for $p$-type structures (Si barriers) in the (100) orientation. 
    In all cases, the results are shown for interface diffusion lengths 
    of 0\,nm, 1\,nm and 2\,nm with arrows denoting the increasing 
    diffusion length.
  }
\end{figure}

Interdiffusion increases the separation of low energy subbands as shown 
in Fig.\ \ref{fig:SubbandSpacing}.  As scattering rates depend on 
subband separation, interdiffusion affects scattering in two possible 
ways:
\begin{enumerate}
\item A \emph{direct} effect due to the change in interface geometry.
\item An \emph{indirect} effect due to the change in subband spacing.
\end{enumerate}
Throughout this section, we adjust the width of QWs to correct the
interdiffusion effect on subband separation.  The calculated change in 
scattering rates is therefore due only to the change in interface
geometry.  In section \ref{scn:GrowthProcess}, the total effect is 
determined by varying interdiffusion without correcting the subband 
separations.

\subsection{(100) $n$-type single QW}
In Si$_{1-x}$Ge$_x$ alloys with $x<85$\%, the conduction band has six
minima near the Brillouin zone edge in the $\Delta$
directions.\cite{article:SSTPaul2004} The valleys are almost parabolic,
and a single band effective mass approximation (EMA) accurately models
electron confinement.\cite{book:Harrison2005}

The $\Delta$ valleys have ellipsoidal equipotential surfaces with their
major axes along the cubic unit cell edge directions. Two separate
effective masses are defined for electrons with wave vectors near the
$\Delta$ minima: the longitudinal mass, $m_l=0.916$ and the transverse
mass, $m_t=0.19$.\cite{article:PRBRieger1993} Quantum confinement in the
$z$ direction in strained (100) systems yields four degenerate
$\Delta_4$ subbands with quantization effective mass, $m_q=m_t$ and
two $\Delta_2$ subbands with $m_q=m_l$. As shown in
fig.\ \ref{fig:InterdiffN100}, the $\Delta_4$ conduction band edge is
at a relatively high energy and it can be assumed to have a negligible
electron population. The following discussion therefore considers only
scattering between $\Delta_2$ subbands and omits $f$-phonon emission
processes.

A more precise double valley EMA models states with $m_q=m_l$ as a
combination of basis states from the two $\Delta_2$ valleys. In $z$
confined systems, the phase difference between reflected basis
components splits the degeneracy of the subbands. We have previously
shown however, that the splitting is quite small for QWs wider than
2--3\,nm and it is therefore omitted in this
work.\cite{article:PRBValavanis2007}

\begin{figure}
 \includegraphics{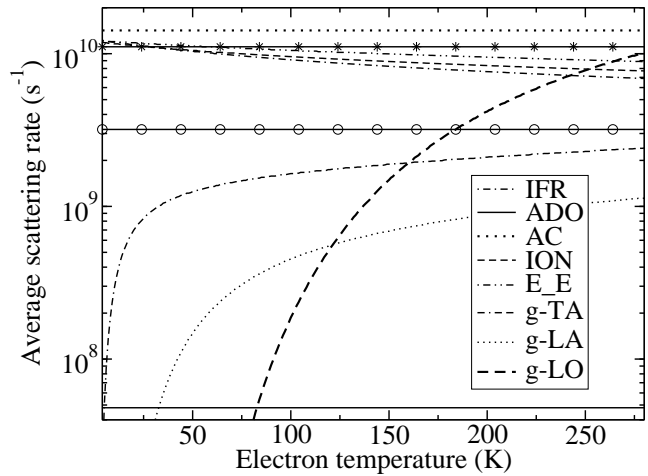}
\caption{\label{fig:TempSweep10meV}Average scattering rates from the
second to first subband in a $n$-type (100) 10\,nm Si QW between two
5\,nm Si$_{0.5}$Ge$_{0.5}$ barriers with $T$=4\,K as a function of
electron temperature. Rates are shown for electron--electron (E\_E),
intravalley acoustic phonon (AC), interface roughness (IFR), ionized
impurity (ION), g-type phonon emission, and alloy disorder (ADO)
processes. ADO is strongly dependent on diffusion length and is shown at
$L$=0\,nm (no symbols), 1\,nm (circles), and 2\,nm (stars). All other
rates are almost independent of interdiffusion and are shown only for
$L$=0\,nm}
\end{figure}

The width of a Si QW between a pair of Si$_{0.5}$Ge$_{0.5}$ barriers (as
in fig.\ \ref{fig:InterdiffN100}) was adjusted for $E_{21}$=10\,meV for a
given diffusion length. Scattering rates were then calculated as a
function of $T_e$ and are shown in fig.\ \ref{fig:TempSweep10meV}.

All mechanisms except alloy disorder scattering were found to be almost
independent of diffusion length and their rates are only plotted for the
abrupt interface system, for simplicity. For all diffusion lengths,
Coulombic interactions and interface roughness rates were relatively
large. As interdiffusion increases however, the alloy disorder rate
increases very rapidly, from 5$\times{}10^{7}$\,s$^{-1}$ for $L$=0\,nm
to 1.1$\times{}10^{10}$\,s$^{-1}$ for $L$=2\,nm. As described
previously, this is due to alloy disorder scattering being permitted in
the well region of diffuse structures, where the electron probability is
large.

The electron temperature dependence of the average alloy disorder and
intravalley acoustic phonon scattering rate is shown to be weak in
fig.\ \ref{fig:TempSweep10meV}, as the scattering potentials depend
neither on the initial wave vector nor explicitly on the electron
temperature. The interface roughness scattering however depends on the
initial wave vector and hence the average rate is affected by the
temperature dependent distribution of electrons in the subband. The
screening of the Coulombic interactions is also affected by electron
temperature, explaining the gradual decrease in average rate. Finally,
the intervalley phonon emissions are affected very strongly by electron
temperature because the subband spacing is lower than $\hbar\omega_0$
for each of the permitted phonons.\cite{article:JAPDollfus1997} The
average scattering rate therefore depends on the number of electrons
with sufficient initial kinetic energy to scatter into a state within
the lower subband.

\begin{figure}
 \includegraphics{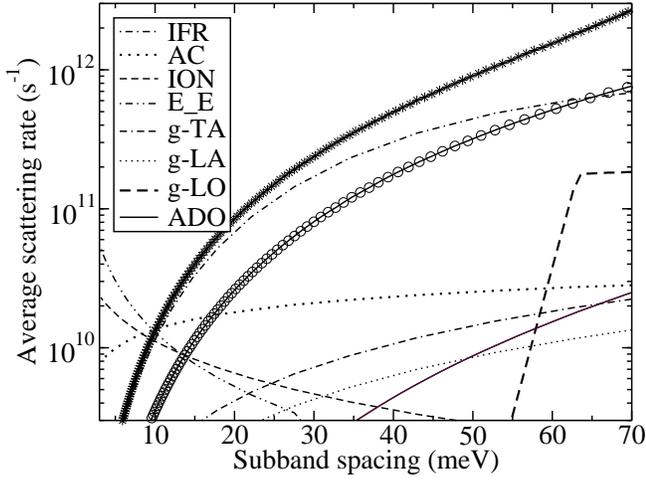}
 \caption{\label{fig:ScattRatesE12}Average scattering rates from second
 to first subband in a (100) $n$-type Si QW as a function of subband
 separation at electron temperature of $T_e$=24\,K. The ADO rate is
 shown at diffusion lengths, $L$=0\,nm (no symbols), 1\,nm (circles), and
 2\,nm (stars). All other rates are almost independent of $L$ and are
 shown only for $L$=2\,nm for simplicity.}
\end{figure}

In pump--probe experiments, which are often used to determine
scattering lifetimes, the carrier temperature, $T_e$ is elevated above
the lattice temperature, $T$ and decays over time.  We have previously
shown however, that good agreement with experimental data is achievable
by assuming $T_e=T+$20\,K.\cite{article:PRBCalifano2007}  In 
fig.\ \ref{fig:ScattRatesE12}, we show the average intersubband 
scattering rate as a function of subband separation at $T=$4\,K, 
$T_e$=24\,K.

For $L$=0\,nm and subband spacing closer than 10\,meV, ionized impurity
and electron--electron scattering dominate, while at spacings between
10\,meV and 55\,meV, interface roughness and intravalley acoustic phonon
scattering are fastest. As the subband spacing becomes comparable to the
energy of the $g$-LO phonons, the emission rate exceeds the intravalley
acoustic phonon scattering rate. Alloy disorder scattering is again
shown to increase significantly with interdiffusion and becomes
dominant for subband spacing above 10\,meV and $L$=2\,nm.

\subsection{(111) $n$-type single QW}
Several important changes are introduced by moving to the (111)
orientation. Electron confinement is now determined using an oblique
cross-section of the ellipsoidal equipotential surface in $\vec{k}$-space,
which is identical in all six $\Delta$ valleys. The quantization
effective mass is\cite{article:JAPRahman2005}
\begin{equation}
 m_q=\frac{3m_lm_t}{2m_l+m_t}=0.26,
\end{equation}
compared with $m_q=m_l$=0.916 for the $\Delta_2$ valleys in (100)
systems. The density-of-states effective mass
is\cite{article:JAPRahman2005}
\begin{equation}
m_d=\sqrt{\frac{m_t(2m_l+m_t)}{3}}=0.36,
\end{equation}
compared with $m_d=m_t$=0.19 for the $\Delta_2$ valleys in (100) systems.

\begin{figure}
 \includegraphics{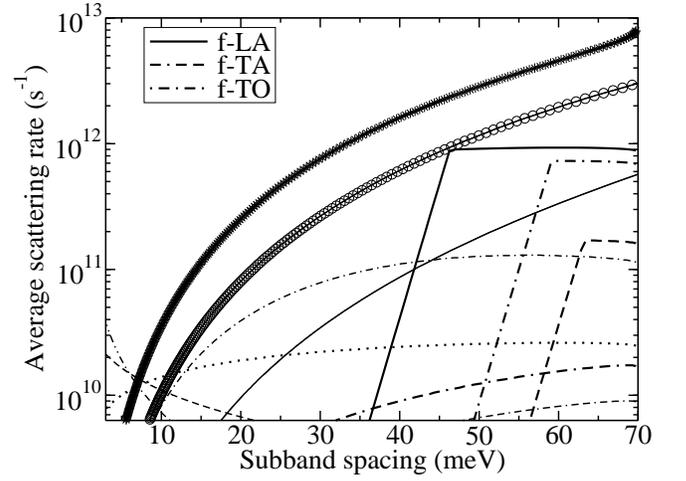}
\caption{\label{fig:ScattRatesDiffT24n111}Average scattering rates from
second to first subband in a (111) $n$-type Si QW as a function of
subband separation with $T_e$=24\,K. All rates from
fig.\ \ref{fig:ScattRatesE12} are shown (using the same legend) as well
as the $f$-phonon rates.}
\end{figure}

The uniaxial strain splitting exhibited in (100) is absent in (111)
systems, while the hydrostatic strain induced shift in conduction band
potential is given by
\begin{equation}
 \Delta_\text{c,av}=\left(\Xi_d+\frac{1}{3}\Xi_u\right)\left(\frac{12c_{44}\varepsilon_\parallel}{c_{11}+2c_{12}+4c_{44}}\right)
\end{equation}
where $c_{ii}$ are elastic constants, $\Xi_{u,d}$ are deformation
potentials, and $\varepsilon_\parallel$ is the in-plane strain as opposed
to
\begin{equation}
 \Delta_\text{c,av}=2\varepsilon_\parallel\left(\Xi_d+\frac{1}{3}\Xi_u\right)\left(\frac{c_{11}-c_{12}}{c_{11}}\right)
\end{equation}
for (100) systems.\cite{article:SSESmirnov2004} For a strain symmetrized
10\,nm Si QW between two 5\,nm Si$_{0.5}$Ge$_{0.5}$ barriers, the
conduction band offset is around 150\,meV, compared with around 380\,meV
for the $\Delta_2$ offset in (100) as shown in
fig.\ \ref{fig:InterdiffN100} As the valley degeneracy is not split, the
population of subbands in each valley is equal and $f$-phonon processes
are no longer negligible. Fig.\ \ref{fig:ScattRatesDiffT24n111} shows the
average intersubband scattering rates in a (111) $n$-type Si QW between
two 5\,nm Si$_{0.5}$Ge$_{0.5}$ barriers with $T_e$=24\,K as a function
of subband separation.

The Coulombic interactions and the intravalley- and $g$-phonon
interactions are almost unchanged compared with the (100) case. From
eqn.~\ref{eqn:IRRateSmooth}, it can be deduced that the reduced
conduction band offset reduces the interface roughness scattering rate
in (111) systems slightly. More significantly, a larger well width is
required to achieve an equivalent subband splitting and hence a smaller
proportion of the wave functions extend over the interface region. The
scattering matrix element is therefore smaller than in the (100) case. 
The alloy disorder scattering is also slightly increased by the change 
in density-of-states effective mass.

The zero-order $f$-phonon emission processes become large at
$E_{12}\gtrsim\hbar\omega_0$ and in structures with abrupt interfaces
represent the dominant mechanisms. $f$-phonon interactions have four
destination valleys and therefore dominate over $g$-phonons, which have
only a single destination.

\subsection{(100) $p$-type single QW}
The band structure for hole transitions is highly nonparabolic and
state contributions from the light hole (LH), heavy hole (HH), and spin
split-off (SO) bands are all significant at nonzero in-plane
wave vectors. The single band EMA is therefore inadequate and the
6$\times$6 $\vec{k}\cdot\vec{p}$ solution described
previously,\cite{article:JAPIkonic2004} was used to account for the
multiband effects.

\begin{figure}
\includegraphics{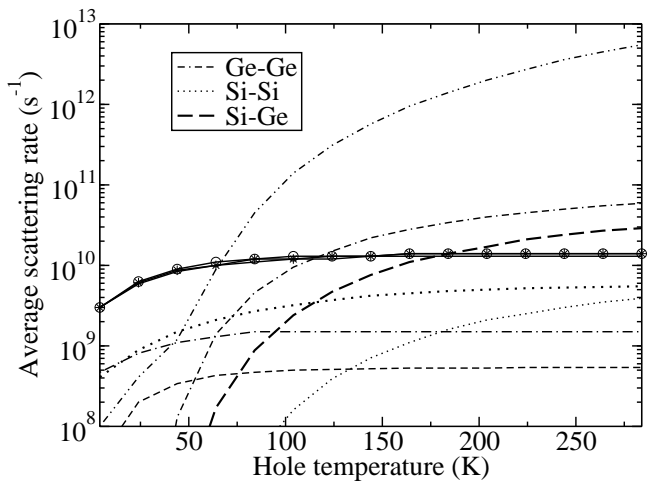}
\caption{\label{fig:TempSweep10meV_HH_HH}Average scattering rate from
second to first HH subband in a $p$-type QW as a function of hole
temperature, with subband separation fixed at 10\,meV. The legend is the
same as previous figures, with the optical phonon branches as labeled.
The alloy disorder scattering rate is shown at diffusion lengths, 
$L$=0\,nm (no symbols), $L$=1\,nm (circles), and $L$=2\,nm (stars).
All other rates are shown at $L$=0\,nm only.}
\end{figure}

As Si/SiGe forms type II heterointerfaces, the well and barrier
compositions of the structure used for $n$-type systems were reversed.
The resulting HH band offset was 350\,meV. The well width was adjusted
to give a 10\,meV separation between the two lowest HH states at
$\vec{k_{\parallel}}=0$. Fig.\ \ref{fig:TempSweep10meV_HH_HH} shows the
average intersubband scattering rates for the system as a function of
hole temperature, $T_h$.

The results differ considerably from those of the $n$-type systems
considered previously. As $T_h$ increases, the hole distribution spreads
over a larger range of in-plane wave vector and HH states acquire a
larger LH contribution. This affects both the scattering matrix element
and the effective density of states, resulting in an increase in all
rates.

The carrier--carrier scattering rate depends on the overlap between
initial and final states for a pair of carriers and is strongly
dependent on the in-plane wave vector of the involved states and hence
on hole distribution over $\vec{k_{\parallel}}$. The $T_h$ dependence is
therefore extremely strong and carrier carrier scattering dominates
above $T_h$=65\,K in this system. Below this temperature, alloy disorder
scattering is dominant.

In contrast with $n$-type systems, alloy disorder scattering in $p$-type
systems is very weakly dependent on interdiffusion. This is because the
magnitudes of wave functions are largest in the center of the QW, where
the Ge fraction is barely affected by the interdiffusion.

\section{\label{scn:GrowthProcess}Growth process tolerance}
We have so far considered the effect of interdiffusion on intersubband
transitions of known energies. Interdiffusion due to growth processes
however changes the subband separation as well as the scattering rates.
There are therefore important implications for fabrication of
intersubband devices such as RTDs and QCLs. The following method uses
both these effects to estimate the tolerance of a device design to
undesired interdiffusion.

\begin{figure}
\includegraphics{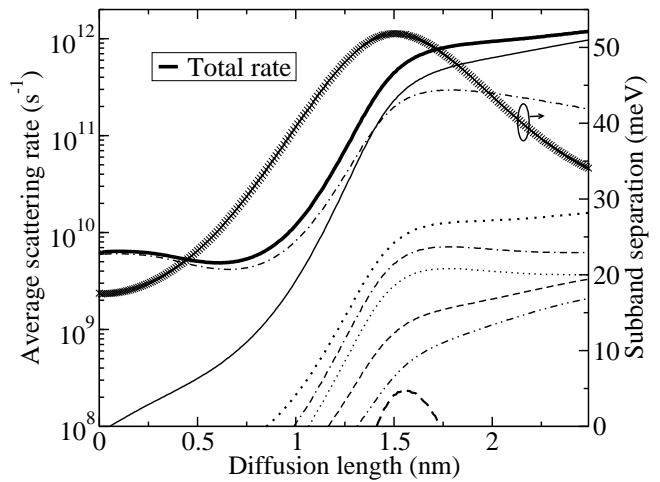}
\caption{\label{fig:ScattRatesT24_vs_diff}Average scattering rates from
second to first subband in a (100) $n$-type double QW with Si wells of
width 5\,nm and 3\,nm separated by a 1\,nm Si$_{0.5}$Ge$_{0.5}$ barrier
as a function of Ge interdiffusion. The legend is the same as that in
fig.\ \ref{fig:ScattRatesE12}. The subband separation as a function of
interdiffusion is overlaid (crosses).}
\end{figure}

A (100) $n$-type double QW system with two coupled Si wells of width
5\,nm and 3\,nm, separated by a 1\,nm Si$_{0.5}$Ge$_{0.5}$ barrier
provides a relatively simple example similar to that used in our
previous pump--probe investigation of intersubband transition lifetimes
in $p$-type materials.\cite{article:PRBCalifano2007} The system was
surrounded by a pair of 5\,nm thick Si$_{0.5}$Ge$_{0.5}$ barriers as
before. For simplicity, the symmetric approximation for interdiffusion
was preserved and the temperatures were fixed at $T$=4\,K and
$T_e$=24\,K. The scattering rates and subband separation are plotted in
fig.\ \ref{fig:ScattRatesT24_vs_diff}.

\begin{figure}
\includegraphics{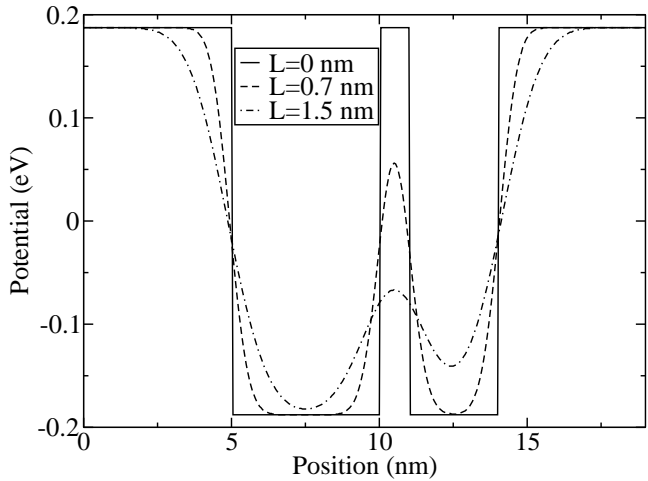}
\caption{\label{fig:TwoWellPotential}$\Delta_2$ conduction band edge
in a two-well heterostructure, nominally comprising 50\% Ge barriers and
Si wells.  The left and right well widths are 5\,nm and 3\,nm 
respectively, and the separating barrier width is 1\,nm.  The potential 
is shown for diffusion lengths of 0\,nm, 0.7\,nm and 1.5\,nm, which 
coincide with the uncoupled, weakly coupled and single well operating 
regimes respectively.}
\end{figure}

In the nominal structure, the layers are defined precisely, restricting
alloy disorder scattering to the barrier regions. The separating barrier
is thick enough for the coupling to be very weak between states. The
subband separation of 17.5\,meV is too low for phonon emission to be
significant, although large enough (and the matrix element small enough)
for Coulombic interactions to be negligible. The dominant rate is
therefore interface roughness scattering, which is also slowed by the
small overlap between states. As the interdiffusion increases, the
device behavior varies and can be characterised by the following
operating regimes, which are illustrated in 
fig.\ \ref{fig:TwoWellPotential}.

\begin{enumerate}
\item \textbf{Uncoupled wells ($L\lesssim0.7$\,nm):} When interdiffusion
is small, extra Ge in the wells increases alloy disorder scattering. The
overlap of states is decreased as the bottom of the barrier becomes
thicker. The other scattering rates therefore decrease slightly.
Conversely, the bottoms of the wells narrow and subband spacing is
increased.

\item \textbf{Weakly coupled wells ($0.7\lesssim{}L\lesssim1.5$\,nm):}
For moderate interdiffusion, the barrier potential is substantially
reduced, while the subband spacing is increased further. As the second
subband minimum approaches the barrier potential, the overlap between
states increases rapidly. All scattering rates therefore begin to rise
extremely rapidly.

\item \textbf{Single well ($L\gtrsim1.5$\,nm):} For large
interdiffusion, the second subband energy exceeds the barrier potential
and the character of the system changes from that of a double QW to a
single QW with a small central perturbation. The perturbation diminishes
as interdiffusion increases and the subband separation decreases
towards the value for a wide, single well. The effect on scattering is
slightly more complex than the previous regimes.

The first-order electron--phonon rates remain almost constant as the
effect of the increased overlap is countered by the decrease in subband
separation. The $g$-LO phonon emission however, decreases sharply due to
its high frequency and zero-order character. Coulombic interactions
continue to increase as the overlap between states increases and the
subband separation decreases. Interface roughness scattering slows as
the barrier potential diminishes and alloy disorder scattering
continues to increase as the Ge content in the single well rises.
\end{enumerate}

The total rate in fig.\ \ref{fig:ScattRatesT24_vs_diff} remains almost
constant at diffusion lengths below 1\,nm, although the subband spacing
varies considerably above 0.25\,nm. A multiple QW device such as a QCL
might therefore operate successfully with 1\,nm interdiffusion if
changes in transition energy are acceptable. Above this tolerance
however, there are catastrophic changes in the device and both
transition energies and lifetimes will be severely affected. This method
may readily be applied to a wide variety of intersubband device designs
as an estimator of robustness.

\section{Conclusion}
We have simulated annealed interfaces to model real systems more
accurately than the rectangular well approximation. The study of carrier
dynamics has thus been extended to systems with diffuse interfaces
(either by design or via the growth process). Intersubband scattering
rates have been investigated as functions of subband separation and
electron temperature in (100) $p$-type and (100) and (111) $n$-type
diffuse QWs.

We have shown that for any given subband separation and electron
temperature, the alloy disorder scattering increases rapidly with
diffusion length in $n$-type systems as Ge content in the well region
increases. The alloy disorder scattering is a relatively minor effect in
systems with abrupt interfaces, while in diffuse systems, it can become
the dominant mechanism. In $p$-type systems however, the effect is
negligible as the Ge content in the center of the well remains large.

In the case of Ge interdiffusion introduced during growth, this
implies that very simple $p$-type system designs may be adjusted to
preserve subband separations and carrier dynamic behavior if the
achievable diffusion length is known. In $n$-type systems however, the
carrier dynamics may be strongly affected even when the design is
adjusted to preserve subband separations.

The robustness of an example double (100) $n$-type QW design was
examined by considering the combined effect of Ge interdiffusion on both
the subband separation and scattering rates. It was shown that a
tolerance of $L<$0.25\,nm preserved the subband separation, while a less
restrictive tolerance of $L<$1\,nm preserved the scattering rates.

As group IV heterostructure epitaxy is less developed than that of
III--V systems, we propose that future designs of complex devices such
as QCLs in the Si/SiGe materials system should be tested for their
robustness using this method before attempting growth.

\begin{acknowledgments}
This work is supported by EPSRC Doctoral Training Allowance funding. The
authors are grateful to Marco Califano and Leon Lever, University of
Leeds for useful discussions.
\end{acknowledgments}
\bibliography{irpaper}
\end{document}